\documentclass[preprints,article,accept,moreauthors,pdftex]{mdpi}

\firstpage{1} 
\pubvolume{1}
\issuenum{1}
\articlenumber{0}
\pubyear{2024}
\copyrightyear{2024}
\datereceived{7 April 2024} 
\daterevised{2 May 2024} % Comment out if no revised date
\dateaccepted{ } 
\datepublished{ } 
\hreflink{https://doi.org/} % If needed use \linebreak
\newcommand{\dm}{{\rm DM}}
\newcommand{\host}{{\rm host}}
\newcommand{\mw}{{\rm MW}}
\newcommand{\igm}{{\rm IGM}}
\newcommand{\halo}{{\rm halo}}

\newcommand{\dd}{{\rm d}}

\newcommand{\reffg}[1]{Figure~\ref{#1}}
\newcommand{\reftb}[1]{Table~\ref{#1}}
\newcommand{\refeq}[1]{Equation~(\ref{#1})}
\newcommand{\refsc}[1]{Section~\ref{#1}}
\Title{Fast Radio Burst Energy Function in the Presence of $\dm_\host$~Variation}

\TitleCitation{Fast Radio Burst Energy Function in the Presence of $\dm_\host$ Variation}

\Author{Ji-Guo Zhang $^{1}$\orcidA{}, Yichao Li $^{1}$*\orcidB{}, Jia-Ming Zou $^{1}$, Ze-Wei Zhao $^{1}$\orcidC{}, Jing-Fei Zhang $^{1}$\orcidD{} and Xin Zhang $^{1,2,3}$*\orcidE{}}
\AuthorNames{Ji-Guo Zhang, Yichao Li, Jia-Ming Zou, Ze-Wei Zhao, Jing-Fei Zhang and Xin Zhang}

\AuthorCitation{Zhang, J.G.; Li, Y.; Zou, J.M.; Zhao, Z.W.; Zhang, J.F.; Zhang, X.}
\address{%
$^{1}$ \quad Key Laboratory of Cosmology and Astrophysics (Liaoning Province) \& Department of Physics, College of Sciences, Northeastern University, Shenyang 110819, China;
zhangjiguo@stumail.neu.edu.cn (J.-G.Z.); neuxiao1@163.com (J.-M.Z.);
zhaozw@stumail.neu.edu.cn (Z.-W.Z.); 
jfzhang@mail.neu.edu.cn (J.-F.Z.)\\
$^{2}$ \quad Key Laboratory of Data Analytics and Optimization for Smart Industry (Ministry of Education), Northeastern University, Shenyang 110819, China\\
$^{3}$ \quad National Frontiers Science Center for Industrial Intelligence and Systems Optimization, 
Northeastern University, Shenyang 110819, China\\}

\corres{Correspondence: liyichao@mail.neu.edu.cn (Y.L.); zhangxin@mail.neu.edu.cn (X.Z.)}

\abstract{Fast radio bursts (FRBs) have been found in great numbers, but the physical mechanism of these sources
is still a mystery. The redshift evolutions of the FRB energy distribution function and the volumetric rate
shed light on the origin of  FRBs. However, such estimations rely on the 
dispersion measurement (DM)--redshift ($z$) relationship. A few FRBs that have been detected recently show large excess DMs
beyond the expectation from the cosmological and Milky Way contributions, which indicates large spread of 
DMs from their host galaxies. In this work, we adopt two lognormal-distributed $\dm_\host$ models and estimate the 
energy function using the non-repeating FRBs selected from the Canadian Hydrogen Intensity Mapping Experiment 
(CHIME)/FRB Catalog 1. By comparing the lognormal-distributed $\dm_\host$ models to a constant $\dm_\host$ model, 
the FRB energy function results are consistent within the measurement uncertainty.
We also estimate the volumetric rate of the non-repeating FRBs in three different redshift bins.
The volumetric rate shows that the trend is consistent with the stellar-mass density redshift evolution. 
Since the lognormal-distributed $\dm_\host$ model increases the measurement errors,
the inference of FRBs tracking the stellar-mass density is nonetheless~undermined.}

\begin{document}

\section{Introduction} \label{sec:intro}
Fast radio bursts (FRBs) are bright, violent flashes of radio emission with durations in the order of milliseconds.     
In 2007, the first FRB event was discovered by~\mbox{\citet{Lorimer:2007qn}} from the archived data of the Parkes telescope in Australia. 
In 2013, four FRB events were discovered by the Green Bank Telescope (GBT)~\cite{thornton2013pop}. In 2017,~\citet{Chatterjee:2017dqg} for the first time confirmed the host galaxy of one FRB, i.e., FRB 20121102A, which was discovered
by the Arecibo radio telescope \citep{Spitler:2014fla}. At this point, $\sim$800 FRB sources have been discovered, with frequencies from hundreds of MHz to several GHz, by advanced radio telescopes~\endnote{\url{https://blinkverse.alkaidos.cn}, accessed on 5 May 2024},  such as
the Canadian Hydrogen Intensity Mapping Experiment
(CHIME)~\endnote{\url{https://www.chime-frb.ca/catalog}, accessed on 5 May 2024} and the Australian Square Kilometre Array Pathfinder
(ASKAP)~\endnote{\url{https://www.atnf.csiro.au/projects/askap/index.html}, accessed on 5 May 2024}, etc. 
Most of the FRBs are non-repeating events, and dozens of sources are repeaters emitting repeating bursts (see ref.~\cite{Hu:2022oie} for a comprehensive review).

There is still an open question of the origin of FRBs (see ref.~\cite{zhang2023solving} for a recent review). A series of theoretical models have been proposed 
to elucidate how FRBs originate \citep{Platts:hiy}. 
{Typically, FRBs can come from strange quark stars. This idea is supported by the $\sim$16-day periodicity of FRB 20180916B, which can be explained by the collapse of the crust of a strange quark star \citep{geng2021rep}}.
Accurate localizations of FRBs within their individual hosts {also} give us hints about their origin. In 2020, the galactic magnetar SGR J1935+2154 \citep{Bochenek:2020zxn,CHIMEFRB:2021srp}  was found to produce an FRB 
that coincided in time with a non-thermal X-ray burst from the magnetar. 
This supports the conjecture that some FRBs may be produced by magnetars.
The repeating FRB 20200120E, localized at the position of a globular cluster in M81 \citep{Bhardwaj:2021xaa}, suggests an old stellar population as the progenitor. This is because globular clusters are usually composed of old stars with low metal content.
By contrast, the repeating FRB 20201124A was localized to a massive, star-forming galaxy \citep{Fong:2021xxj} at a redshift of $z = 0.098$~\citep{Nimmo:2021ntn}.  There is no consensus in determining the type of FRB hosts. Nevertheless, the redshift of an FRB can be inferred from its unique host galaxy, which enables valuable cosmological applications as the localized FRBs accumulate in great numbers \citep{Gao:2014iva,Zhou:2014yta,Zhao:2020ole,Qiu:2021cww,Wu:2022dgy,Zhao:2022bpd,Yang:2022ftm,Wang:2022ami,Zhang:2023gye,Wei:2023avr}; see refs.~\cite{Bhandari:2021thi,Xiao:2021omr} for recent reviews. 
For more information on FRB hosts, refer to the comprehensive catalog of the 23 localized FRBs with secure hosts provided by ref.~\cite{Gordon:2023cgw}. 
The limited number of localized events to date, however, hinders our ability to fully understand the origin of FRBs.

The FRB energy function \citep{Zhang:2020ass,Luo:2020wfx,James:2021oep} is also an effective way to constrain the origin models of FRBs. The energy functions allow us to study the redshift evolution of the volumetric rate of FRBs. To achieve this, the intensity distribution function can also be studied \citep{Li:2016qbl}.
%Study the luminosity function or energy function of FRB, which can tell us the volumetric FRB rates%. 
If FRBs originate from young stellar populations, the volumetric FRB rate should rise with increasing redshift as the density of the cosmic star formation rate increases towards a higher redshift (up to $z$ $\sim$ 2). %Please ensure meaning has been retained. 
Conversely, if FRB progenitors are old, such as white dwarfs,
neutron stars, or black holes, the volumetric rate should follow the evolution of stellar mass, 
which is barely evolved at low redshift.
%{showing little variation with increasing redshift.} 
Ref.~\cite{Hashimoto:2020acj} found no 
significant redshift evolution in the number density of non-repeating FRB sources, which is %Please ensure meaning has been retained. 
consistent with the stellar-mass evolution in the universe.
In their recent research, ref.~\cite{Hashimoto:2022llm} reported that old progenitors like neutron stars and black holes are more likely to be the progenitors of non-repeating FRBs.

%They use the observed dispersion measurement method to derive the redshift of each FRB and adopt 
%a constant $\dm_\host$ = 50 ${\rm pc}\,{\rm cm}^{-3}$. 
Recently, a few FRBs have shown larger excess DMs than what is typically expected from
the cosmological and Milky Way contributions 
\citep{Spitler:2014fla,Chatterjee:2017dqg,
Hardy:2017djf,Tendulkar:2017vuq,chittidi:2021}. 
In particular,~\citet{Niu:2021bnl} reported the detection of FRB 190520B 
with $\dm_{\host} \approx 903^{+72}_{-111}\,{\rm pc}\,{\rm cm}^{-3}$, 
which is almost an order of magnitude higher than the average $\dm_\host$ of the FRBs discovered so far. The uncertainties from $\dm_\host$ can inevitably skew the inferred redshifts of unlocalized FRBs when employing the DM--$z$ relationship for estimation \citep{Walker:2018qmw,James:2022dcx}, which poses challenges to revealing their underlying engine and potential use as cosmological probes. Therefore, it is crucial to investigate the impact of $\dm_\host$ uncertainty on the redshift inference of FRBs and the associated population properties. In this work, we study the effect of $\dm_\host$ uncertainty on the FRB energy function estimation as well as the redshift evolution of the FRB event rate. Note that in this work, the {\it Planck} 2018 $\Lambda$ cold dark matter model is adopted as a fiducial model, with the best-fit cosmological parameters $H_0=67.36$ km s$^{-1}$ Mpc$^{-1}$, $\Omega_{\rm b}=0.0493$, $\Omega_{\rm m}=0.3153$, and $\Omega_{\Lambda}=0.6847$.

This paper is organized as follows:
In \refsc{sec:data}, we describe the FRB catalog as well as the selection criteria 
used in this work.
The Bayesian framework used for the redshift estimation is described in \refsc{sec:method}. 
In \refsc{Results}, we present the energy functions and volumetric rates of non-repeating FRB sources along with their redshift evolution. The conclusions are presented in \refsc{ccc}.

\section{Data}\label{sec:data}

\subsection{FRB Catalog}\label{sec:frbcat}

We use the first release of the CHIME/FRB catalog~\citep{CHIMEFRB:2021srp},
which contains 474 non-repeating bursts and 62 repeating bursts from $18$ repeaters.
The $536$ burst events were detected from 2018 July 25 to 2019 July 1 
over an effective survey duration of $214.8$ days. 
In order to minimize the selection effects, a number of criteria are suggested
\citep{Shin:2022crt,Hashimoto:2022llm}.

\begin{enumerate}
\item Events with {\tt bonsai\_snr}$<10$ are rejected, where {\tt bonsai\_snr} 
is the signal-to-noise ratio (S/N) recorded in the catalog. 
Ref.~\cite{Shin:2022crt} suggests a S/N cut of $12$ since the signal below ${\rm S/N} =12$ may be 
misclassified as radio-frequency interference (RFI).
In this work, we use bursts with S/Ns over $10$, which maintains a meaningful number of FRB samples for the 
statistical analysis \citep{Hashimoto:2022llm}.
\item Events with ${\rm DM}_{\rm obs} < 1.5\times {\rm max}\left({\rm DM}_{\rm NE2001}, {\rm DM}_{\rm YMW16}\right)$ are rejected to ensure the extragalactic origin of the events.
${\rm DM}_{\rm obs}$ is the measured DM; ${\rm DM}_{\rm NE2001}$ and 
${\rm DM}_{\rm YMW16}$ are the DMs of the Milky Way estimated according to 
the NE2001 model \citep{Cordes:2002wz} and the YMW16 model \citep{Yao_2017}, 
respectively.
\item Events with $\lg\left(\tau_{\rm scat}/{\rm ms}\right) > 0.8$ are rejected, 
where $\tau_{\rm scat}$ is the scattering timescale. Note that $\lg$ is equivalent to $\log_{10}$ in this work. 
\item Events with $\lg\left(F_\nu/{\rm Jy\,ms}\right) < 0.5$ are rejected, where $F_\nu$
is the fluence of the burst. 
\item Events detected in the side lobes of the telescope's primary beam are rejected.
\end{enumerate}

After applying the selection criteria, we have $176$ FRB events selected, including $12$~repeaters. 

Previous analysis with mock data indicated that a significant
fraction of FRBs were missed by the CHIME detection algorithm, i.e., only 39,638 out of
84,697 injected mock events were detected \citep{Shin:2022crt}. The total number of events ($N_{\rm FRB}$) needs to be scaled from the observed number of events  ($N_{\rm obs}$) according to the detection fraction:
\begin{equation}\label{eq:selnum}
N_{\rm FRB} = N_{\rm obs} \times \frac{84,697}{39,638}.
\end{equation}

% N_{\rm obs} 未解释
In addition, the fraction of missed events also depends on the properties of the FRB signals.
Longer scattering times or lower fluencies result in a higher number of missed events. Following ref.~\cite{CHIMEFRB:2021srp}, the relationship between the observed and intrinsic data distributions is described by
\begin{equation}
P(\vartheta) = P_{\rm obs}(\vartheta) \times s(\vartheta)^{-1},
\end{equation}
where $P(\vartheta)$ and $P_{\rm obs}(\vartheta)$ represent the intrinsic and observed distributions of the
FRB property $\vartheta$, respectively. The symbol $s(\vartheta)$ is the selection function as a function
of different FRB properties. 
The properties considered for deriving the selection function include
the dispersion measure (${\rm DM}_{\rm obs}$), scattering timescale ($\tau_{\rm scat}$), 
intrinsic duration ($w_{\rm int}$), and fluence ($F_{\nu}$); see ref.~\cite{CHIMEFRB:2021srp} for details.
We adopt the best-fit selection functions in ref.~\cite{Hashimoto:2022llm}:
\begin{eqnarray}
s(\widehat{{\rm DM}}_{\rm obs}) &=& - 0.7707 \Big(\lg\widehat{{\rm DM}}_{\rm obs}\Big)^2
+ 4.5601\Big(\lg\widehat{{\rm DM}}_{\rm obs}\Big) - 5.6291, \label{eq:seldm}\\
s(\widehat{\tau}_{\rm scat})    &=& - 0.2922 \Big(\lg\widehat{\tau}_{\rm scat}\Big)^2 
- 1.0196\Big(\lg\widehat{\tau}_{\rm scat}\Big) + 1.4592, \label{eq:seltau}\\
s(\widehat{w}_{\rm int})        &=& - 0.0785 \Big(\lg\widehat{w}_{\rm int}\Big)^2 
- 0.5435 \Big(\lg\widehat{w}_{\rm int}\Big) + 0.9574, \label{eq:selw}\\
\lg s(\widehat{F}_{\nu})  &=& 1.7173 \Big(1 - \exp\big(-2.0348 \lg \widehat{F}_{\nu}\big)\Big) - 1.7173, \label{eq:selF}
\end{eqnarray}
where $\widehat{{\rm DM}}_{\rm obs}=\Big(\frac{{\rm DM}_{\rm obs}}{{\rm pc\,cm^{-3}}}\Big)$,
$\widehat{\tau}_{\rm scat}=\Big(\frac{\tau_{\rm scat}}{\rm ms}\Big)$,
$\widehat{w}_{\rm int}=\Big(\frac{w_{\rm int}}{{\rm ms}}\Big)$,
and $\widehat{F}_\nu=\Big(\frac{F_\nu}{\rm Jy\,ms}\Big)$.

% including 145 FRB sources (one-off FRBs)

%We use the catalog of contains a total of 536 bursts from 492 sources, including 474 non-repeaters and 18 repeaters from the first release of CHIME/FRB Catalog 1. Its observation frequency ranges from 300 MHz to 8 GHz and dispersion measure (DM) range from 100 to 2600. However, these data are selected by us, DM of data needs to meet the conditions of \( D M>100   \mathrm{pc}   \mathrm{cm}^{-3} \) and \( D M<1200  \mathrm{pc}  \mathrm{cm}^{-3} \), to make sure the sources are extragalactic and the Macquart relation \citep{Macquart:2020lln} between the dispersion measure and redshift (DM-z) is effective.

\subsection{Galaxy Catalog}

%In the analysis of gravitational waves (GW), if the redshift information of the GW source could be obtained by identifying the electromagnetic (EM) counterparts, it is called the bright siren method. While for the GW events without EM counterparts, the statistical analysis of the GW event related to the galaxy catalog can also be applied in obtaining the redshift information. This kind of GW standard sirens is known as the dark sirens \citep{DelPozzo:2011vcw} (see Refs.~\citep{} for recent works). We develop a statistical method to measure the  redshift of FRBs using galaxy catalogs\citep{Zhao:2022yiv}. Thanks to the advanced radio equipment like the Canadian Hydrogen Intensity Mapping Experiment (CHIME)~\endnote{https://www.chime-frb.ca/catalog}, the Australian Square Kilometre Array Pathfinder (ASKAP)~\endnote{https://www.atnf.csiro.au/projects/askap/index.html}, Square Kilometer Array (SKA). More FRBs and their redshifts can be found through these instruments in the future. In this work, the first question to be answered is how to obtain the redshift of FRBs more accurately.% 

In order to evaluate the redshifts of the unlocalized FRBs, we follow the
method developed in ref.~\cite{Zhao:2022yiv}, which actually employs the 
dark siren method in gravitational wave cosmology \citep{DelPozzo:2011vcw,Wang:2022oou,Song:2022siz}.
%{Jin:2023zhi,Jin:2022tdf,Jin:2022qnj,Wang:2021srv,Zhang:2019ylr,Zhang:2019ple,Zhao:2019gyk,Qiu:2021cww}. 
It assumes that an FRB is always located in a galaxy and that the redshift of the 
FRB can be statistically estimated by associating the FRB event with its 
potential host galaxies according to an underlying galaxy catalog.

\begin{figure}[H]%[H]
%    \centering
    \includegraphics[width=0.6\textwidth]{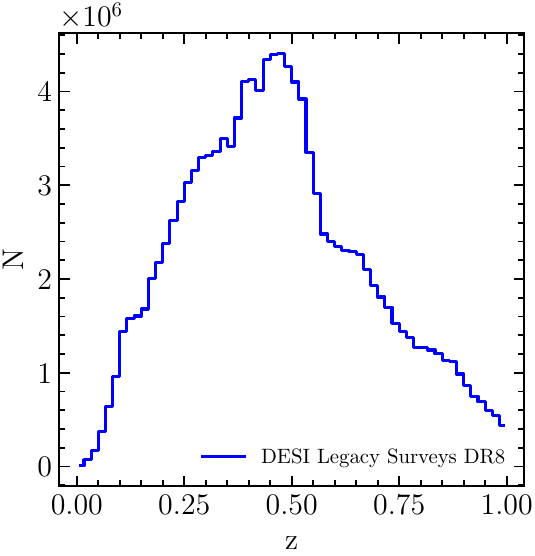}
    \caption{The redshift distribution of DESI Legacy Surveys DR8 galaxy catalog.}\label{fig:galaxy_fig}
\end{figure}

\begin{figure}[H]%[H]
%    \centering
    \includegraphics[width=0.98\textwidth]{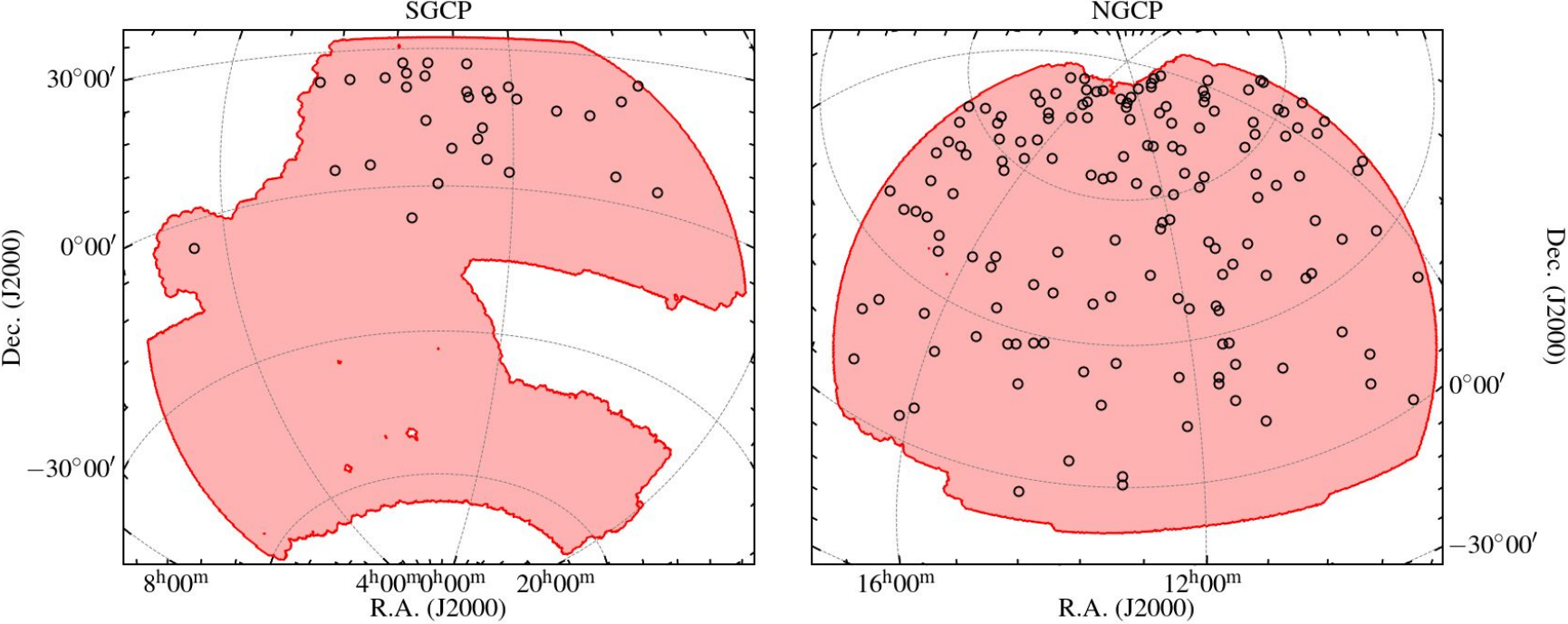}
    \caption{The sky locations of $145$ selected FRB sources (one-off FRBs) within the region covered by the DR8 galaxies. The FRBs are shown with the black circles, and the footprint of the galaxy sample used in this work is shown with the red area. The galaxies within the SGCP and the NGCP are shown in the left and right panels, respectively. 
    }\label{fig:footprint}
\end{figure}

In this work, we adopt the galaxy catalog from 
Dark Energy Spectroscopic Instrument (DESI) Legacy Surveys.
The Legacy Surveys combine three imaging projects of different telescopes, i.e., 
the Beijing--Arizona Sky Survey (BASS)~\citep{zou2017project}, 
the Dark Energy Camera Legacy Survey (DECaLS) \citep{DES:2015wtr},
and the Mayall z-band Legacy Survey (MzLS) \citep{2016AAS...22831702S}, 
covering about $14000\,{\rm deg}^2$ of the northern
hemisphere and producing the target catalog for the DESI survey;
for an overview of the Legacy Surveys, see ref.~\cite{DESI:2018ymu}.
We use the galaxy sample from the eighth public data release of
the Legacy Surveys, i.e., Data Release $8$ (DR8)~\citep{BOSS:2011dbi}.
The spectroscopic redshift of the galaxy sample is substituted for the photometric 
redshift, if available, in accordance with the sample selection process in~ref.~\cite{Yang:2020eeb}. In total, 
the galaxy sample incorporates 129.35 million galaxies, among which 2.1 million galaxies have spectroscopic redshifts. The redshift distribution of the DR8 galaxies is shown in Figure~\ref{fig:galaxy_fig}.
{Finally, 145 one-off FRBs are selected, as they reside within the region covered by the galaxy catalog,
%The sky locations of 145 selected FRB sources (one-off FRBs) are shown in \reffg{fig:footprint} with the black circles.
and their sky locations} are shown in \reffg{fig:footprint} with the black circles.
The footprint of the galaxy sample is illustrated with the red area, where the galaxies residing in the south galactic cap (SGCP) and the north galactic cap (NGCP) are shown in the left and right panels, respectively.

\unskip
% with apparent magnitudes limited 

% {The SDSS-DR8 optical photometric survey over redshift range $0 \leq z \leq 1.1$, with redshift distribution taken from~\citet{Sheldon:2011fm}, which is illustrated in \reffg{}. The code is publicly available~\endnote{\url{https://github.com/dcurl47/probwts}}.} 
% They are selected with r $\textless$ 21.8, 

%We use the catalogGC o f galaxy clusters\citep{Yang:2020eeb} based on the DESI Legacy Imaging Surveys DR8\citep{Yang:2004an} expansion which has a redshift ranging from 0 to 1. The sky coverage of galaxies is divided into the northern Galaxy cap (NGC) and the southern Galaxy cap (SGC), NGC covers about 8580 square degrees and has 59.6 million galaxies, SGC covers about 9673 square degrees with 69.75 million galaxies. It contains the basic properties of the distribution of richness, halo mass, and luminosity.

\section{Methods}\label{sec:method}

%\subsection{redshift estimation}
\subsection{Bayesian Framework}
We adopt a Bayesian data analysis scheme to measure the FRBs' redshifts. 
The Bayesian inference relates the probability density functions 
(PDFs) involving data and parameters:
\begin{equation}
P(\vartheta | x) \propto P(\vartheta) P(x | \vartheta),
\end{equation}
where $P(x | \vartheta)$ is the likelihood function of the data given the
model parameters, and $P(\vartheta | x)$ is the posterior PDF, i.e.,
the PDF of the parameters given the data set.
In this work, we shall estimate the posterior PDF of the FRBs' redshifts $z$ 
given the measurement set of DM:
\begin{equation}
P(z|{\rm DM}) \propto P(z) P({\rm DM}|z),
\end{equation}
where $P(\dm | z)$ represents the likelihood function of the measured DM
given the parameter set.
The measured DM is the combination of several components:
\begin{equation}
\dm = \dm_{\mw} + \dm_{\halo} + \dm_{\igm} + \frac{\dm_{\host}}{1 + z},
\end{equation}
where the contributions are from the interstellar medium of the Milky Way ($\dm_{\mw}$),
the ionized gas in the local halo ($\dm_{\halo}$), the intergalactic medium ($\dm_{\igm}$),
and the FRB host galaxy ($\dm_{\host}$). 

% $\dm_{\mw}$ and $\dm_{\halo}$ can be subtracted according to the current models.
The variable %Please ensure meaning has been retained. 
$\dm_{\mw}$ can be subtracted according to the current models.
The CHIME/FRB catalog provides the dispersion measure with the Milky Way contribution 
subtracted using the NE2001~\citep{Cordes:2002wz} or YMW16 models \citep{Yao_2017}. We test both of the models and find no significant difference 
in the final estimation. In the following analysis, only the results with the YMW16 model
are presented.

The precise contribution of the local halo to DMs is uncertain. 
%with estimates of order 30-245 ${\rm {pc~cm^{-3}}}$
Ref.~\cite{Yamasaki:2019htx} provides the prediction of $\dm_{\halo}$
with a mean value of $43\,{\rm pc}\,{\rm cm}^{-3}$ and a full
range of $30$--$245\,{\rm pc}\,{\rm cm}^{-3}$.
We adopt the mean value of $43\,{\rm pc}\,{\rm cm}^{-3}$ in the following
analysis.

After subtracting $\dm_{\mw}$ and $\dm_{\halo}$ for the total $\dm$, 
the measurement likelihood function is written as
\begin{eqnarray}\label{eq:likeli}
P(\dm|z) = \int &\dd&\dm_{\host}\,\dd\,\dm_{\igm}\,
          P(\dm|\dm_{\host},\dm_{\igm}, z) \nonumber \\
         &\times& P(\dm_{\host}|z) P(\dm_{\igm}| z),
\end{eqnarray}
where $P(\dm_{\host}|z)$ and $P(\dm_{\igm}|z)$ are the 
{likelihood} functions of $\dm_{\host}$ and $\dm_{\igm}$, respectively,
and the integration represents the marginalization  of
the $\dm_{\host}$ and $\dm_{\igm}$ {likelihood} function.

\subsection{$P(\dm_{\igm}|z)$}
The DM contribution from the IGM ($\dm_{\igm}$) can be explained 
as the dispersion induced when an FRB is emitted at a random point in the
universe of redshift $z$ and propagates to $z=0$.
The average value of $\dm_{\igm}$ at redshift $z$ is given by the 
integration of the free electron number density $n_{\rm e}$ along the line of sight:
\begin{equation}
\langle \dm_{\igm} \rangle = \int_0^z \dd z^\prime \frac{n_{\rm e}(z^\prime)}{1+z^\prime}
\left(\frac{1}{1+z^\prime}\frac{c}{H_0}\frac{1}{E(z^\prime)}\right),
\end{equation}
In this work, we consider the standard flat $\Lambda$CDM model
$E(z)=\sqrt{\Omega_{\rm m}(1+z)^3 + \Omega_{\Lambda}}$.
Assuming the universe is fully ionized at $z\lesssim3$, the free electron
number density equals the total electron number density: 
\begin{equation}
n_{\rm e}(z) = f_{\igm}\bar{\rho}_{{\rm b},0}(1+z)^3 
\left( \frac{Y_{\rm H}\chi_{{\rm e},{\rm H}}(z)}{m_{\rm p}} 
     + 2\frac{Y_{\rm He}\chi_{{\rm e},{\rm He}}(z)}{4m_{\rm p}}
\right),
\end{equation}
where $Y_{\rm H}\sim3/4$ and $Y_{\rm He}\sim1/4$ denote the primordial mass 
fractions of hydrogen and helium, respectively. 
The ionization fractions $\chi_{{\rm e}, {\rm H}}(z)$ and
$\chi_{{\rm e}, {\rm He}}(z)$ for hydrogen and helium are both
set to unity at the redshift of $z\lesssim3$ \citep{Fan:2006dp,McQuinn:2008am}.
The symbol %Please ensure meaning has been retained. 
$\bar{\rho}_{{\rm b},0} = 3H_0^2\Omega_{\rm b}/8\pi G$ is the 
comoving cosmological baryon density at the current epoch, $m_{\rm p}$ 
is the mass of a proton, and $f_{\igm} \approx 0.83$ represents the fraction 
of the free electrons in the IGM \citep{Deng:2013aga}.

The $\dm_{\igm}$ deviation from $\langle\dm_{\igm}\rangle$ is 
expected to follow the normal distribution. Thus, the {likelihood}
function is expressed as
\begin{equation}\label{eq:ligm}
P(\dm_{\igm} | z ) = \frac{1}{\mathcal{N}_{{\rm IGM}}}\exp\left( -\frac{1}{2}\frac{\left( {\rm DM}_{\rm IGM} -\langle\mathrm{DM}_{\mathrm{IGM}}\rangle \right)^2}{\sigma^2_{{\rm IGM}}} \right),
\end{equation}
where $\mathcal{N}_{\igm}=\sigma_{\igm} \sqrt{2\pi}$ is the normalization factor,
and $\sigma_{\igm}$ is fitted in a power-law form~\citep{Qiang:2021bwb}:
 \begin{equation}\label{eq:sigigm}
 	\sigma_{\rm{IGM}}=173.8~z^{0.4}~\rm{pc}~\rm{cm}^{-3}.
 \end{equation}
 
\subsection{$P(\dm_{\host}|z)$}
The major uncertainty of the FRB redshift measurement comes from the 
variation of the DM contribution from the FRB's host galaxy.
A few FRBs show large excess DMs beyond the expectation from the
the cosmological and Milky Way contributions~\mbox{\citep{Spitler:2014fla,Chatterjee:2017dqg,
Hardy:2017djf,Tendulkar:2017vuq}.} 
Recently,~\citet{Niu:2021bnl} reported the detection of FRB 190520B 
with $\dm_{\host} \approx 903^{+72}_{-111}\,{\rm pc}\,{\rm cm}^{-3}$, 
which is almost an order
of magnitude higher than the average $\dm_\host$ of the FRBs discovered so far. Broadly speaking, the large spread of the $\dm_\host$ can be modeled 
using a lognormal distribution, and the corresponding {likelihood}
function is expressed as
\begin{equation}\label{eq:lhost}
P(\dm_\host | z) = \frac{1}{{\mathcal N}_{\host}}
\exp\left( -\frac{1}{2} \frac{\left( \ln x - \mu \right)^2}{\sigma^2_{\host} }\right),
\end{equation}
where $x ={{\rm DM}_{\rm host}}\big/{{\rm pc}\,{\rm cm}^{-3}}$, 
${\mathcal N}_{\host}= x \sigma_{\host} \sqrt{2 \pi}$ is the
normalization factor, and $\mu$ and $\sigma_\host$ are both the lognormal distribution 
parameters. Using a cosmological magnetohydrodynamical simulation,
\citet{Mo:2022qxz} proposed a detailed analysis of the distribution 
of $\dm_\host$ for various FRB population models. In this work, we adopt the
fitting result of
\begin{equation}\label{eq:mo20}
	\mu = \ln (63.55),\,\,\sigma_\host=1.25
\end{equation}
from~\citet{Mo:2022qxz} (hereafter referred to as the Mo22 model).
In addition,~\mbox{\citet{Zhang:2020mgq}} (hereafter referred to as the Zhang20 model) provided another fitting result:
\begin{equation}\label{eq:zhang20}
\mu = \ln \left(32.97 (1 + z) ^{0.84}\right),\,\,\sigma_\host=1.248.
\end{equation}

We compare the differences in results obtained using such two $\dm_\host$ distribution models (expressed in Equations~(\ref{eq:mo20}) and (\ref{eq:zhang20})), as well as use the model of assuming a constant $\dm_\host=50\,{\rm pc}\,{\rm cm}^{-3}$. 

\subsection{The Posterior Distribution of the FRB Redshifts}

\begin{figure}[H]%[H]
    \centering
    \includegraphics[width=0.96\textwidth]{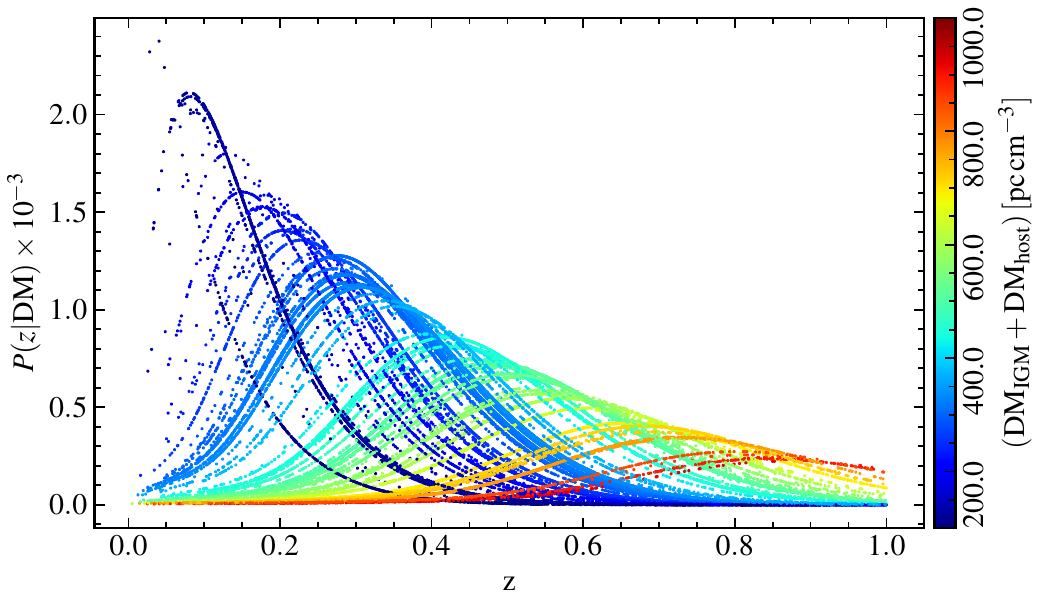}
    \caption{The redshift posterior probability distribution of each FRB event.
    Each curve represents the posterior probability of one FRB using the
    DESI Legacy Surveys DR8 galaxy redshift sample that located in 
    the sky area determined by \refeq{eq:beam}.
    The colors of the lines indicate the values of 
    ${\rm DM}_{\rm IGM}+{\rm DM}_{\rm host}$.
    }\label{fig:likeli}
\end{figure}

The DM measurement {likelihood} function is expressed as
\begin{equation}\label{eq:ldm}
P(\dm | \dm_\host, \dm_\igm, z) = \frac{1}{\mathcal{N}} 
\exp\left(-\frac{1}{2}\frac{(\dm - \Theta)^2}{\sigma^2_{\dm}}\right),
\end{equation}
where $\mathcal{N} = \sigma_\dm \sqrt{2\pi}$ is the normalization factor,
$\sigma_\dm$ is the measurement uncertainty, and 
$\Theta=\dm_\host + \dm_\igm + \dm_\mw + \dm_\halo$ represents the
DM's theoretical value.
Substituting Equations (\ref{eq:ligm}), (\ref{eq:lhost}), and (\ref{eq:ldm}) 
into \refeq{eq:likeli}, we can estimate the posterior probability at a given redshift.
Assuming the FRBs are always located in the galaxies, we shall use the 
redshifts of the galaxy catalog, i.e., the DESI Legacy Surveys DR8 galaxy catalog, 
as the prior distribution. 
For a given FRB, we utilize the redshifts of the selected galaxies in the DR8 catalog with their celestial coordinates $\alpha$ (right ascension) and $\delta$ (declination) satisfying the following criteria:
\begin{eqnarray}\label{eq:beam}
| \alpha_{\rm gal} - \alpha_{\rm FRB} | < \theta_\alpha \times \cos(\delta_{\rm FRB}), \,\,\,
| \delta_{\rm gal} - \delta_{\rm FRB} | < \theta_\delta,
\end{eqnarray}
where $\alpha_{\rm gal}$ and $\delta_{\rm gal}$ are the coordinates of the galaxies, while $\alpha_{\rm FRB}$ and $\delta_{\rm FRB}$ are the coordinates of the FRB. $\theta_\alpha$ and $\theta_\delta$ are the $68\%$ confidence 
pointing errors of the CHIME beam in the right ascension and declination
directions, respectively.
The estimated redshift posterior probabilities are shown in \reffg{fig:likeli}.
%{Each curve, composed of discrete points, represents the posterior probability for a single FRB. The probability is calculated with the DESI Legacy Surveys DR8 galaxies' redshift sample residing in the sky area determined by \refeq{eq:beam}.}
For each FRB event, the posterior probability is estimated at the redshift of each potential host galaxy, 
which is selected from the DESI Legacy Surveys DR8 galaxies sample residing in the sky area determined by \refeq{eq:beam}.
% (Each {discontinuous} curve is in shape of points of the posterior probability calculated at the redshifts of all potential hosts.)
% {Note that each discrete probability is calculated at the redshifts of all potential hosts, and with more redshift candidates, these points tend to form a continuous curve.}
The colors indicate the value of $\dm_\igm + \dm_\host$ of the FRB event.
There is a clear trend of FRBs with larger DMs having their
posterior probability distribution peaking at higher redshifts, which is consistent with the Macquart relation \citep{Macquart:2020}.
However, the posterior probability also shows a wide range of 
distribution, indicating the large uncertainty of the redshift estimation.
Such large uncertainty is primarily due to the large scattering of $\dm_\igm$ (see Equation (\ref{eq:sigigm})). Additionally, the prior distribution of the galaxy catalog from DESI Legacy Surveys DR8 introduces uncertainty, particularly at high redshifts, where the photometric redshifts of galaxies are less accurate.

\subsection{Energy Function}

The FRB fluence ($F_\nu$) is converted to rest-frame isotropic radio energy ($E$) for each FRB via
\citep{Macquart:2018jlq}:
\begin{equation}\label{eq:rho}
E = \frac{4\pi d_{\rm L}^2}{(1+z)^{2+\alpha_{\rm F}}} F_{\nu} \Delta\nu,
\end{equation}
where $d_{\rm L}$ is the luminosity distance to the FRB, $\alpha_{\rm F}$ is the spectrum index of 
the FRB's power-law spectrum across frequencies (note that $\alpha_{\rm F}$ differs from $\alpha_{\rm FRB}$ in \refeq{eq:beam}),
%~\endnote{{Some FRBs with narrow spectra do not confirm an available power-law index, such as the repeating FRB 20201124A discussed in~\citet{Zhou:2022nnh}. According to~\citet{Pleunis:2021qow}, most of one-off CHIME FRBs' spectra are broadband and can be well described by a power-law function. Thus, we assume that all one-off CHIME FRBs used in this work have a power-law index and take $\alpha_{\rm F}=-1.50$ from~\citet{CHIMEFRB:2021srp}. Note that $\alpha_{\rm F}$ differs from $\alpha_{\rm FRB}$ in \refeq{eq:beam}.}}, 
and $\Delta \nu$ is the burst bandwidth. The burst bandwidth is calculated by 
{\tt high\_freq}$-${\tt low\_freq} in the CHIME/FRB Catalog 1, 
where {\tt high\_freq} and {\tt low\_freq} represent the upper and lower bands, respectively,
of the detection at full-width tenth-maximum (FWTM) \citep{CHIMEFRB:2021srp}.
The spectrum index term,  $(1+z)^{-\alpha_{\rm F}}$, represents the bandwidth $k$-correction 
since the event emitted its radiation in a different band than that in which it was observed.
It is known that there are narrow-band FRBs, such as the repeating FRB 20201124A discussed in~\citet{Zhou:2022nnh},
and the spectrum index for the non-power-law spectrum is tricky to define.
The CHIME/FRB Catalog 1 provided the spectrum index for both the broad- and narrow-band FRBs
but used an additional spectrum-running parameter, reshaping the spectra to match different bandwidths. 
As long as the $k$-correction is negligible for the narrow-band spectrum, \refeq{eq:rho} is available 
for both broad- and narrow-band FRBs.
In addition, the narrow-band FRBs are mostly repeating FRBs~\cite{Pleunis:2021qow}.
This work focuses on the non-repeating FRBs, which are mostly broad-band ones.
We assume a constant power-law index and take $\alpha_{\rm F}=-1.5$ from~\citet{CHIMEFRB:2021srp}. 

The FRB energy function represents the number density of FRB events as a function of energy. 
{To estimate this function, we adopt the $V_{\rm max}$ method \citep{Schmidt:1968kn,avni:1980}, which defines $V_{\rm max}$ as the flux-limited maximum volume within which 
the FRB event could still be detected~\citep{Schmidt:1968kn,avni:1980}:}
\begin{equation}
	V_{\rm max} = \frac{4\pi}{3}\left(\chi_{\rm max}^3 - \chi_{\rm min}^3\right),
\end{equation}
where $\chi_{\rm min}$ and $\chi_{\rm max}$ are the comoving distances at the 
minimum and maximum redshifts, respectively. We adopt $z_{\rm min}=0.05$ in this work, and 
$z_{\rm max}$ is estimated according to the fluence of the FRB:
\begin{equation}
	F_{\nu} = \frac{E(1+z_{\rm max})^{2+\alpha_{\rm F}}}{4\pi d_{{\rm L},z_{\rm max}}^2\Delta \nu } 
	> 10^{0.5}\,{\rm Jy}\,{\rm ms}.
\end{equation}

{The number density for a single FRB event is defined as the inverse of $V_{\rm max}$,
%the volume corresponding to the the maximum comoving distance where the FRB can be detected
and the number density per unit of time is expressed as}
%The number density of each FRB event observation (${\rho}_{\rm obs}$) per unit time is estimated,
\begin{equation}\label{eq:rhoobs}
{\rho}_{\rm obs} = \frac{1}{V_{\rm max} f_{\rm sky} \left( t_{\rm obs}/(1+z) \right)},
\end{equation}
where $f_{\rm sky}=3\times10^{-3}$ is the fraction of the sky covered by the
CHIME's field of view, $t_{\rm obs}=0.59\,{\rm yr}$ is the survey time for
the CHIME/FRB Catalog 1, and the factor of $1+z$ converts the survey time to the rest frame.
%Then the number density estimated with {\refeq{eq:rhoobs}} 
{Furthermore, ${\rho}_{\rm obs}$} needs to be corrected for the selection
effect, as mentioned in \refsc{sec:frbcat}. Considering the selection function of
Equations~(\ref{eq:seldm})--(\ref{eq:selF}), the corrected number density is
\citep{Hashimoto:2022llm}
\begin{equation}
\rho_{\rm corr} = \frac{1}{\mathcal{N}_{s}}W\rho_{\rm obs},
\end{equation}
where $W=\big( s(\dm_{\rm obs}) s(\tau_{\rm scat}) s(w_{\rm int}) s(F_{\nu}) \big)^{-1}$. 
$\mathcal{N}_s = \sum_i W_i / {N_{\rm FRB}}$ is the normalization
factor, where $i$ denotes the $i$-th FRB event, and $N_{\rm FRB}$ is the corrected total number defined in \refeq{eq:selnum}.

% We divide the full energy range occupied by the FRB detection into a number of 
% energy bins ($\phi_j$) in the logarithmic scale and sum $\rho_{\rm corr}$ within
% each energy bin,
% \begin{equation}
% \phi_{j} = \frac{1}{\Delta_j \lg E}\sum_i \rho_{{\rm corr}, i},
% \end{equation}
% where $\Delta_j \lg E$ is the $j$-th energy bin size.
We divide the full energy range occupied by the FRB detection into a number of energy bins in the logarithmic scale and sum $\rho_{\rm corr}$ within each energy bin; then, the FRB energy function is
\begin{equation}
\phi_{j} = \frac{1}{\Delta_j \lg E}\sum_i \rho_{{\rm corr}, i},
\end{equation}
where $\Delta_j \lg E$ is the $j$-th energy bin size.

We perform a Monte Carlo (MC) simulation with $10000$ realizations of the
redshift sample following the posterior probability distribution of 
\refeq{eq:likeli}. The energy function is estimated using each of the realizations.
The uncertainty is evaluated via the the standard deviation.

%a “normalized” Schechter type is selected as described in the Equation (10),
%\begin{equation}
%P(E)dE=\rho\frac{1}{E_{\text {char }}}\left(\frac{E}{E_{\text {char }}}\right)^\gamma \exp \left[-\frac{E}{E_{\text {char }}}\right]dE
%\end{equation}
%where$ \rho $ is a normalization coefficient, E is the radio energy, $E_{\text {pivot }}$ is the pivot energy of FRBs. $E_{\text {char }}$ is the cut-off energy of FRBs. $\gamma$ is the power-law index.~\cite{Shin:2022crt} used the Schechter function to infer a characteristic energy cut-off of $\rm{E_{char}}=2.38_{-1.64}^{+5.35}$×$10^{41}$, power-law index of $\gamma=-1.3_{-0.4}^{+0.7}$ and FRBs event rate of $\left.\left[7.3_{-3.8}^{+8.8} \text { (stat.) }{ }_{-1.8}^{+2.0} \text { (sys. }\right)\right] \times 10^4$Gpc$^{-3}$ year$^{-1}$, the above three parameters are fitted to our energy distribution, as is shown in Fig.~\ref{fig:2}. We infer a characteristic energy cut-off of $\rm{E_{char}}$=1.01×$10^{41}$ erg, a power-law index of ${\gamma}$= -0.86 and a volumetric rate 5.75×$10^{4}$ of bursts Gpc$^{-3}$ year$^{-1}$above a pivot energy of $10^{39}$ erg. This result is basically consistent with their research results.
%

% \begin{figure}[H]%
% %\begin{minipage}[t]{0.49\textwidth}
% \centering

% \caption{}\label{fig:efMo}                             
% \end{figure}
% %\end{minipage} \hfill
% %\begin{minipage}[t]{0.49\textwidth}

\section{Results and Discussion}\label{Results}

\begin{figure}[H]%[!htbp]
%\centering
%\includegraphics[width=\textwidth]{plot_EF_onceoff145_ymw16_fixdm.pdf}
\includegraphics[width=0.96\textwidth]
{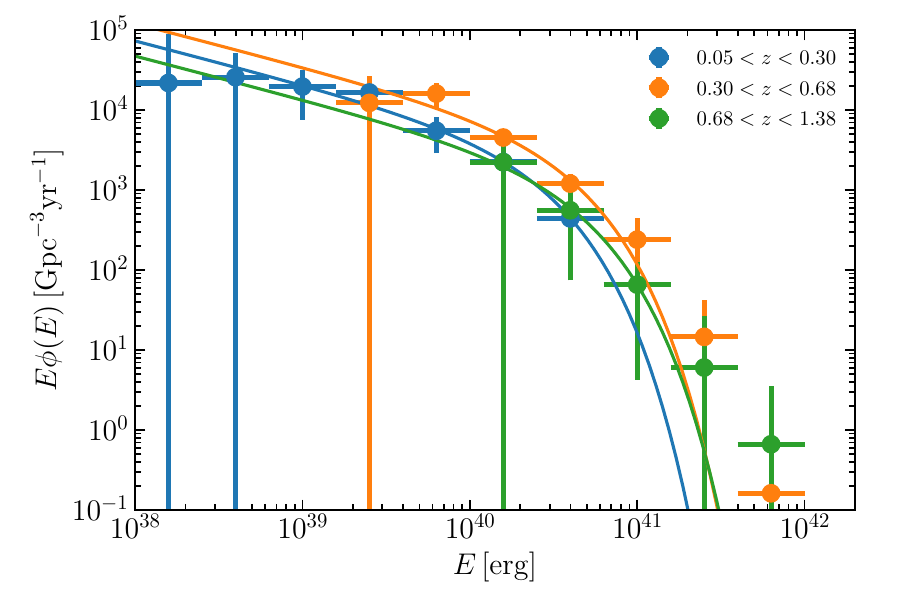}
\includegraphics[width=0.96\textwidth]{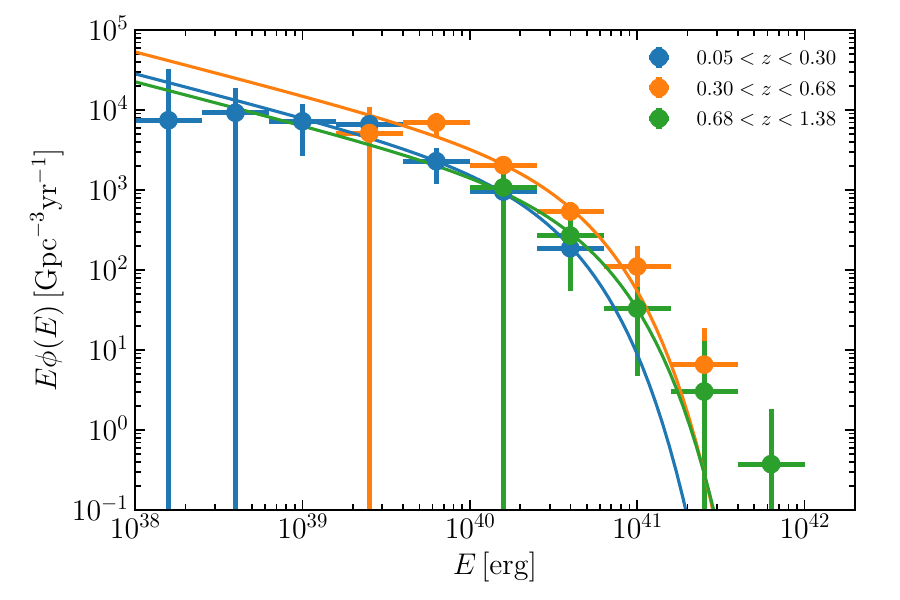}
\caption{The energy functions of non-repeating CHIME FRB sources using the Mo22 model (upper panel) and the Zhang20 model (bottom panel). In each panel, the results for three redshift bins are shown with three different colors, and the best-fit Schechter functions are shown with solid lines.}\label{fig:efMoZhang}                             
%\end{minipage}
\end{figure}

Figure \ref{fig:efMoZhang} shows the energy functions estimated using the 
CHIME/FRB Catalog 1. The energy functions are estimated within three redshift bins,
i.e., \mbox{$0.05<z \leqslant 0.30$, $0.30<z \leqslant 0.68$,} and $0.68<z \leqslant 1.38$ (following ref.~\citep{Hashimoto:2022llm}).
The estimated energy distribution functions of different redshift bins are shown 
in different colors.
The results with different $\dm_\host$ models, i.e., the Mo22 model and the Zhang20 model, are shown in the upper and bottom panels of Figure \ref{fig:efMoZhang}, respectively.

The FRB energy distribution is modeled with a Schechter function \citep{1976ApJ...203..297S}:
\begin{align}\label{eq:ef}
\phi(\lg E)\,{\rm d}\lg E = \phi^\star \left(\frac{E}{E^\star}\right)^{\gamma + 1}
\exp\left(-\frac{E}{E^\star}\right) {\rm d} \lg E,
\end{align}
where $\phi^\star$ is the normalization factor, $\gamma + 1$ is the faint-end slope, and $E^\star$ is the break energy of the Schechter function. 
The energy distribution functions are fitted to the measurements using 
the public package {\tt emcee}~\endnote{\url{https://emcee.readthedocs.io/en/stable/index.html}, accessed on 5 May 2024}
\citep{Foreman:2012any}.
We use $\phi^\star$, $E^\star$, and $\gamma$ as free parameters for
the first redshift bin, i.e., $0.05 < z \leqslant 0.30$.
Due to the lack of FRB data in the higher redshift bins, 
we only use $\phi^\star$ and $E^\star$ as the free parameters in
the remaining two redshift bins and fix $\gamma$ to the best-fit values of the
first redshift bin.

The best-fit energy distribution functions at each redshift bin are shown with the smooth curves in Figure \ref{fig:efMoZhang}. The energy functions estimated in each redshift bin are mutually consistent. There is no significant redshift evolution for using either 
the Mo22 model or the Zhang20 model. The best-fit values of parameters in 
\refeq{eq:ef} are listed in \reftb{tab:bestfit}. The volumetric rate of the FRBs ($\Phi_{\rm FRB}$) is estimated by integrating the energy function within the energy range available for each redshift bin, which spans $10^{37}$--$10^{43}~\rm erg~\rm s^{-1}$ according to ref.~\cite{Luo:2020wfx}.

\begin{figure}[H]%[!htbp]
%\centering
\includegraphics[width=0.98\textwidth]{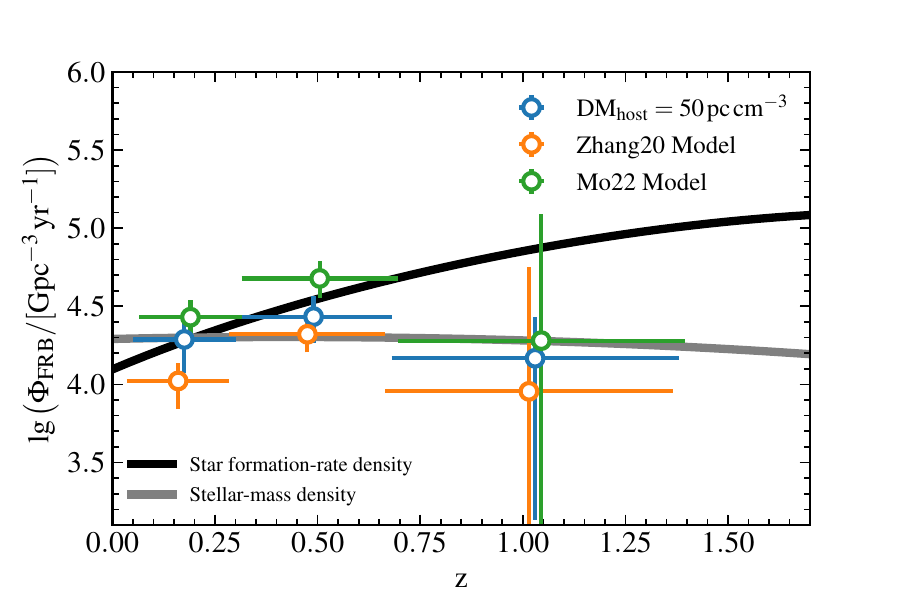}
\caption{The volumetric rate of CHIME non-repeating FRBs as a function
of redshift. The horizontal errors represent the redshift bin width, and the vertical errors indicate the estimation
uncertainty; estimates are based on 10000 iterations %Please ensure meaning has been retained. 
of the MC simulation. 
The results from the constant $\dm_\host$ model, the Zhang20 model, and the Mo22 model are shown with different colors. The solid black and gray curves show the cases of the star formation rate and the stellar-mass density, respectively, which are both estimated using the fitting functions in ref.~\cite{Hashimoto:2020acj}.
%The blue errorbars shows the result by assuming constant $\dm_\host$, while the orange and green errorbars show the results using $\dm_\host$ followingZhang20 Model and Mo22 Model, respectively.
}\label{fig:frbrate}
\end{figure}
% the relevant analysis from the relevant analysis from 
%\begin{align}\label{eq:frbrate}
%\Phi = \int_{E_{\rm min}}^{E_{\rm max}} \phi(E) {\rm d} \ln E,
%\end{align}

% The corresponding volumetric rates estimated using the best-fit energy functions
% are listed in the last column of \reftb{tab:bestfit} and also shown in \reffg{fig:frbrate}.
% In \reffg{fig:frbrate}, the blue error bars show the FRB volumetric rates estimated
% by assuming a constant $\dm_\host=50\,{\rm pc}\,{\rm cm}^{-3}$, while the 
% orange and green error bars show the results using $\dm_\host$ from the Zhang20  and Mo22 models, respectively.
% The horizontal error bars indicate the redshift bin width, while the vertical error bars show the standard deviation. 
% The solid black and gray curves show the cases of the star formation rate and the stellar-mass density, respectively, which are both estimated using the fitting functions in Ref.~\cite{Hashimoto:2020acj}.

The corresponding volumetric rates estimated using the best-fit energy functions  are listed in the last column of \reftb{tab:bestfit} and are also shown in \reffg{fig:frbrate}.
The figure illustrates the FRB volumetric rates for different $\dm_\host$ models. It also includes solid lines of the star formation rate and the stellar-mass density, each estimated using the fitting functions in ref.~\cite{Hashimoto:2020acj}.
Both of the curves are normalized for their amplitudes at redshift $z=0.2$ to the same value as the FRB volumetric rates estimated using the constant $\dm_\host$.

With the constant $\dm_\host$ assumption, the FRB volumetric rate shows the same
trend as the stellar-mass density, which is consistent with the previous analysis
\citep{Hashimoto:2022llm}. By releasing the constant $\dm_\host$ assumption, 
the uncertainty increases, especially for the high redshift bin. 
Within the estimated error, there is no significant difference between 
using and not using the constant $\dm_\host$ assumption. 
However, it can be seen that the variation of $\dm_\host$ weakens the conclusion that the volumetric rate is consistent with the case of stellar-mass density. It is expected that future, much larger FRB and galaxy samples could greatly improve the measurement and help draw a more solid conclusion.
%\unskip

\begin{table}[H]
%\begin{center}
%\centering
%\setlength{\tabcolsep}{3mm}
%\renewcommand{\arraystretch}{1.5}
\caption{The best-fit parameters (i.e., $\phi^\star$, $E^\star$, and $\gamma$) in the Schechter function for the FRB energy function using the CHIME/FRB Catalog 1 with different redshift bins and $\dm_{\host}$ models, and the estimated volumetric rates of the FRBs ($\Phi_{\rm FRB}$) are shown in the last column. Here, $E^{*}$ is in units of erg,  while $\phi^{*}$ and $\Phi_{\rm FRB}$ are both in units of $\rm Gpc^{-3}~\rm yr^{-1}$.}\label{tab:bestfit}

\begin{adjustwidth}{-\extralength}{0cm}
\centering %% If there is a figure in wide page, please release command \centering

\begin{tabular}{m{4.3cm}<{\centering}m{3cm}<{\centering}m{3cm}<{\centering}m{3cm}<{\centering}m{3cm}<{\centering}} \toprule
& \boldmath{$\lg E^\star$} & \boldmath{$\lg \phi^\star$} & \boldmath{$\gamma$} & \boldmath{$\lg \Phi_{\rm FRB}$} \\ \midrule
\multicolumn{5}{c}{$0.05 < z \leqslant 0.30$} \\ \midrule
Constant $\dm_\host$  & $40.000_{-0.363}^{+0.428}$ & $3.994_{-0.626}^{+0.254}$ & $-1.046_{-0.373}^{+0.516}$ & $4.277_{-0.301}^{+0.139}$ \\
Zhang20 Model         & $40.219_{-0.378}^{+0.592}$ & $3.418_{-0.877}^{+0.358}$ & $-1.382_{-0.224}^{+0.572}$ & $4.035_{-0.192}^{+0.125}$ \\
Mo22 Model            & $40.225_{-0.382}^{+0.589}$ & $3.783_{-0.882}^{+0.369}$ & $-1.427_{-0.238}^{+0.537}$ & $4.438_{-0.181}^{+0.121}$ \\
\midrule
\multicolumn{5}{c}{$0.30 < z \leqslant 0.68$} \\ \midrule
Constant $\dm_\host$  & $40.239_{-0.119}^{+0.169}$ & $4.097_{-0.222}^{+0.164}$ & -- & $4.434_{-0.167}^{+0.129}$ \\
Zhang20 Model         & $40.503_{-0.150}^{+0.177}$ & $3.370_{-0.226}^{+0.207}$ & -- & $4.321_{-0.116}^{+0.113}$ \\
Mo22 Model            & $40.496_{-0.153}^{+0.183}$ & $3.729_{-0.235}^{+0.211}$ & -- & $4.678_{-0.123}^{+0.111}$ \\
\midrule
\multicolumn{5}{c}{$0.68 < z \leqslant 1.38$} \\ \midrule
Constant $\dm_\host$  & $40.408_{-0.451}^{+0.216}$ & $3.772_{-0.847}^{+0.336}$ & -- & $4.167_{-1.033}^{+0.263}$ \\
Zhang20 Model         & $40.553_{-1.549}^{+0.461}$ & $2.970_{-0.470}^{+1.640}$ & -- & $3.956_{-1.869}^{+0.794}$ \\
Mo22 Model            & $40.547_{-1.519}^{+0.534}$ & $3.295_{-0.003}^{+2.204}$ & -- & $4.280_{-1.992}^{+0.813}$ \\ \bottomrule
\end{tabular}

\end{adjustwidth}
%\end{center}
\end{table}

%It means that the variation of $\dm_\host$ do not have significant affect on the FRB volumetric rate prediction. Such measurements can be improved in the future with even large FRB catalog and deeper galaxy catalog.  

\section{Conclusions}\label{ccc}

In this work, we estimate the energy function and the volumetric rate of the non-repeating
FRBs using the CHIME/FRB Catalog 1. We follow the FRB selection criteria as used in refs.~\citep{Shin:2022crt,Hashimoto:2022llm}. In the meantime, we follow the
Bayesian framework data analysis scheme developed in ref.~\cite{Zhao:2022yiv} and adopt the galaxy catalog from DESI Legacy Surveys to estimate redshifts of the unlocalized FRBs.

% We also consider different $\dm_\host$ models, including a {constant-${\rm DM}_{{\rm host}}$} model of assuming $\dm_\host = 50\,{\rm pc}\,{\rm cm}^{-3}$, as well as a couple of lognormal distribution models, which are constrained using the magnetohydrodynamical simulation. Based on this simulation,~\citet{Mo:2022qxz} and~\citet{Zhang:2020mgq} both provided the fitting functions (see Equations~(\ref{eq:mo20}) and (\ref{eq:zhang20})) referred to as the Mo22 model and Zhang20 models, respectively. The FRB energy function is estimated with each of the $\dm_\host$ models.

We also consider different $\dm_\host$ models, including a constant-${\rm DM}_{{\rm host}}$ model of assuming $\dm_\host = 50\,{\rm pc}\,{\rm cm}^{-3}$ as well as a couple of lognormal distribution models: namely, the Mo22 model and the Zhang20 model. The FRB energy function is estimated with each of the $\dm_\host$ models.

The Schechter-function-like energy function model is considered and fitted to the
measurements using the non-repeating FRBs from the CHIME/FRB Catalog 1. 
% The best-fit values of {the} parameters are summarized in \reftb{tab:bestfit}.
We do not find a significant difference between using the constant $\dm_\host$ model
and the two lognormal $\dm_\host$ models (i.e., the Mo22 model and the Zhang20 model).

We also estimate the FRB volumetric rates according to the best-fit energy 
distribution function and compare the trends of redshift evolution with the
star formation rate density and the stellar-mass density. 
We find that with the lognormal $\dm_\host$ models, the uncertainties
increase. The trend of redshift evolution is consistent with the 
stellar-mass density for the constant $\dm_\host$ model and the lognormal 
$\dm_\host$ models.
However, since the lognormal-distributed $\dm_\host$ model increases the measurement errors, the inference of FRBs tracking the stellar-mass density is nonetheless undermined.
The measurement can be further improved in the future by using a larger FRB catalog and/or a deeper galaxy survey catalog.

\authorcontributions{Conceptualization, Y.L. and X.Z.; methodology, J.-G.Z.; software, Y.L. and J.-G.Z.; validation, J.-G.Z and J.-M.Z.; formal analysis, J.-F.Z.; investigation, Z.-W.Z.; writing---original draft preparation, Y.L.; writing---review and editing, J.-G.Z; supervision, J.-F.Z. and X.Z.; project administration, X.Z. All authors have read and agreed to the published version of the manuscript.}

\funding{\textls[-18]{This research was funded by the National SKA Program of China (grant Nos. 2022SKA0110200} and 2022SKA0110203) and the National Natural Science Foundation of China (grant Nos. 11975072, 11875102, and 11835009).}

\institutionalreview{Not applicable.}

\informedconsent{Not applicable.}

\dataavailability{The code that supports this work is publicly available at \url{https://github.com/YichaoLi/frb_efunc} (accessed on 5 May 2024).}

\acknowledgments{We thank Chenhui Niu and Yuhao Zhu for helpful discussions and suggestions.
We are grateful for the support from the National SKA Program of China (grant Nos. 2022SKA0110200 and 2022SKA0110203) and the National Natural Science Foundation of China (grant Nos. 11975072, 11875102, and 11835009).}

\conflictsofinterest{The authors declare no conflicts of interest.}

\begin{adjustwidth}{-\extralength}{0cm}
\printendnotes[custom] 
\reftitle{References}
%\bibliography{FRBefunc}

\PublishersNote{}

\end{adjustwidth}
\end{document}